\documentstyle[psfig]{mn}

\newif\ifAMStwofonts
\AMStwofontstrue

\newcommand{\be}{\begin{equation}}
\newcommand{\ee}{\end{equation}}
\newcommand{\ba}{\begin{eqnarray}}
\newcommand{\ea}{\end{eqnarray}}
\newcommand{\brr}{\begin{array}}
\newcommand{\err}{\end{array}}
\newcommand{\bc}{\begin{center}}
\newcommand{\ec}{\end{center}}

\newcommand{\mincir}{\raise
  -2.truept\hbox{\rlap{\hbox{$\sim$}}\raise5.truept \hbox{$<$}\ }}
\newcommand{\magcir}{\raise
  -2.truept\hbox{\rlap{\hbox{$\sim$}}\raise5.truept \hbox{$>$}\ }}
\newcommand{\siml}{\raise
  -2.truept\hbox{\rlap{\hbox{$\sim$}}\raise5.truept \hbox{$<$}\ }}
\newcommand{\simg}{\raise
  -2.truept\hbox{\rlap{\hbox{$\sim$}}\raise5.truept \hbox{$>$}\ }}

\ifoldfss
\ifCUPmtlplainloaded \else
\NewTextAlphabet{textbfit} {cmbxti10} {}
\NewTextAlphabet{textbfss} {cmssbx10} {}
\NewMathAlphabet{mathbfit} {cmbxti10} {} 
\NewMathAlphabet{mathbfss} {cmssbx10} {} 
\fi
\ifAMStwofonts
\ifCUPmtlplainloaded \else
\NewSymbolFont{upmath} {eurm10}
\NewSymbolFont{AMSa} {msam10}
\NewMathSymbol{\upi}     {0}{upmath}{19}
\NewMathSymbol{\umu}     {0}{upmath}{16}
\NewMathSymbol{\upartial}{0}{upmath}{40}
\NewMathSymbol{\leqslant}{3}{AMSa}{36}
\NewMathSymbol{\geqslant}{3}{AMSa}{3E}

\fi
\fi
\fi 

\ifnfssone
\newmathalphabet{\mathit}
\addtoversion{normal}{\mathit}{cmr}{m}{it}
\addtoversion{bold}{\mathit}{cmr}{bx}{it}
\newmathalphabet{\mathbfit} 
\addtoversion{normal}{\mathbfit}{cmr}{bx}{it}
\addtoversion{bold}{\mathbfit}{cmr}{bx}{it}
\newmathalphabet{\mathbfss} 
\addtoversion{normal}{\mathbfss}{cmss}{bx}{n}
\addtoversion{bold}{\mathbfss}{cmss}{bx}{n}
\ifAMStwofonts
\ifCUPmtlplainloaded \else
\UseAMStwoboldmath
\makeatletter
\new@mathgroup\upmath@group
\define@mathgroup\mv@normal\upmath@group{eur}{m}{n}
\define@mathgroup\mv@bold\upmath@group{eur}{b}{n}
\edef\UPM{\hexnumber\upmath@group}
\new@mathgroup\amsa@group
\define@mathgroup\mv@normal\amsa@group{msa}{m}{n}
\define@mathgroup\mv@bold\amsa@group{msa}{m}{n}
\edef\AMSa{\hexnumber\amsa@group}
\makeatother
\mathchardef\upi="0\UPM19
\mathchardef\umu="0\UPM16
\mathchardef\upartial="0\UPM40
\mathchardef\leqslant="3\AMSa36
\mathchardef\geqslant="3\AMSa3E
\fi
\fi
\fi 

\ifnfsstwo
\DeclareMathAlphabet{\mathbfit}{OT1}{cmr}{bx}{it}
\SetMathAlphabet\mathbfit{bold}{OT1}{cmr}{bx}{it}
\DeclareMathAlphabet{\mathbfss}{OT1}{cmss}{bx}{n}
\SetMathAlphabet\mathbfss{bold}{OT1}{cmss}{bx}{n}
\ifAMStwofonts
\ifCUPmtlplainloaded \else
\DeclareSymbolFont{UPM}{U}{eur}{m}{n}
\SetSymbolFont{UPM}{bold}{U}{eur}{b}{n}
\DeclareSymbolFont{AMSa}{U}{msa}{m}{n}
\DeclareMathSymbol{\upi}{0}{UPM}{"19}
\DeclareMathSymbol{\umu}{0}{UPM}{"16}
\DeclareMathSymbol{\upartial}{0}{UPM}{"40}
\DeclareMathSymbol{\leqslant}{3}{AMSa}{"36}
\DeclareMathSymbol{\geqslant}{3}{AMSa}{"3E}
\fi
\fi
\fi 

\ifCUPmtlplainloaded \else
\ifAMStwofonts \else 
\def\upi{\pi}
\def\umu{\mu}
\def\upartial{\partial}
\fi
\fi

\title[Simulating the metal enrichment of the ICM] {Simulating the
  metal enrichment of the intra--cluster medium} \author[Tornatore et
  al.]  {L. Tornatore$^{1}$, S. Borgani$^{1,2}$, F. Matteucci$^{1}$,
  S. Recchi$^{3}$, P. Tozzi$^{4}$ \\~\\ $^1$ Dipartimento di
  Astronomia dell'Universit\`a di Trieste, via Tiepolo 11, I-34131
  Trieste, Italy (borgani,matteucci,tornatore@ts.astro.it)\\ $^2$ INFN --
  National Institute for Nuclear Physics, Trieste, Italy\\ $^3$
  Max-Planck-Institut f\"ur Astrophysik, Karl-Schwarzschild-Strasse 1,
  D-85740 Garching bei München, Germany\\ (recchi@mpa-garching.mpg.de)\\
  $^4$ INAF, Osservatorio Astronomico di Trieste, via Tiepolo 11,
  I-34131 Trieste, Italy (tozzi@ts.astro.it)}

\pubyear{2003}

\begin{document}
\label{firstpage}
\maketitle

\begin{abstract}
We present results from Tree+SPH simulations of a galaxy cluster,
aimed at studying the metal enrichment of the intra--cluster medium
(ICM). The simulation code includes a fairly advanced treatment of
star formation, as well as the release of energy feedback and detailed
yields from both type-II and type-Ia supernovae, also accurately
accounting for the lifetimes of different stellar populations. We
perform simulations of a cluster with virial mass $\simeq 3.9\times
10^{14}M_\odot$, to investigate the effect of varying the feedback
strength and the stellar initial mass function (IMF). Although most of
the models are able to produce acceptable amounts of Fe mass, we find
that the profiles of the iron abundance are always steeper than
observed. The [O/Fe] ratio is found to be sub--solar for a Salpeter
IMF, with [O/Fe]$\simeq -0.2$ at $R\magcir 0.1R_{200}$, whereas
increasing to super-solar values in central regions, as a result of
recent star formation. Using a top--heavier IMF gives a larger [O/Fe]
over the whole cluster, at variance with observations. On the other
hand, the adoption of a variable IMF, which becomes top-heavier at
$z>2$, provides a roughly solar [O/Fe] ratio. Our results indicate that
our simulations still lack a feedback mechanism which should quench
star formation at low redshift and transport metals away from the star
forming regions.
\end{abstract}

\begin{keywords}
Cosmology: Theory -- Galaxies: Intergalactic Medium -- Methods:
Numerical -- $X$--Rays: Galaxies: Clusters
\end{keywords}

\section{Introduction}
Measurements of the content and distribution of metals in the
intra--cluster medium (ICM) provides invaluable insights on the
interplay between the evolution of diffuse cosmic baryons and the past
history of star formation. If supernovae (SN) had played a significant
role in altering the thermal status of the ICM, then they should have
left their imprint also on its metal content (e.g., Renzini 2003).
The increasing capability of X--ray telescopes to perform spatially
resolved spectroscopic studies of the ICM have opened in the last
years the possibility of quantifying the metal content of
clusters. Observations from ASCA (e.g., Baumgartner et al. 2003),
Beppo--SAX (e.g., De Grandi et al. 2003), Chandra (e.g., Ettori et
al. 2002; Blanton et al. 2003) and XMM--Newton (e.g., Gastaldello \&
Molendi 2002; Matsushita et al. 2003) have revealed that abundance
gradients are quite common in clusters. Besides their distribution,
the relative abundance of different metal species gives information on
the role played by type-Ia and II SN in the ICM enrichment (e.g.,
Matteucci \& Vettolani 1988; Pipino et al. 2002; Finoguenov et
al. 2002; Portinari et al. 2003).  Finally, the lack of any
significant evolution of the Fe abundance at least out to $z\sim1$
(Tozzi et al. 2003) demonstrates that the process of enrichment has
been completed already at quite large lookback times.  While the
production of metals is connected to the activity of star formation,
their distribution may be determined by different physical effects,
such as ram--pressure stripping of metal-rich gas from merging
galaxies or galactic winds powered by SN explosions and AGN activity
(e.g., Gnedin 1998). Whatever the mechanism for metal transport and
diffusion is, modelling the ICM enrichment requires a careful
description of the yields and of the lifetimes of the different
stellar populations associated to the different SN types (e.g.,
Matteucci 2001, and references therein).  In this respect N--body
simulations provide the ideal tool to describe in detail how metals
are produced within galaxies and distributed during the hierarchical
assembly of a cluster. Besides semi--analytical approaches (e.g. De
Lucia, Kauffman \& White 2003), attempts to include, within
hydrodynamical simulations, star formation, SN energy feedback and
metal enrichment from type--Ia and II SN, have been pursued by
different authors (Aguirre et al. 2001; Lia, Portinari \& Carraro
2002; Valdarnini 2002; Kawata \& Gibson 2003; Kobayashi 2003; Tissera
\& Scannapieco 2003). It is however clear that such approaches rely on
the capability of the numerical codes to provide a physically sound
description of the relevant ``sub--grid'' processes.

In this {\em Letter} we present the first results from our
hydrodynamical simulations of clusters based on the implementation of
chemical enrichment in the GADGET code (Springel, Yoshida \& White
2001). Our chemo-dynamical version of GADGET combines the rather
advanced treatment of star formation and SN feedback, proposed by
Springel \& Hernquist (2003a, SH03 hereafter), to a careful description
of the role of type--Ia and II SN in releasing metal--enriched gas into
the diffuse medium. In the following, when expressing the
ICM metal abundances in solar units, we assume the photospheric
abundance provided by Grevesse \& Sauval (1998).

\begin{figure*}
\vspace{-5mm}
\centerline{\hbox{
\psfig{file=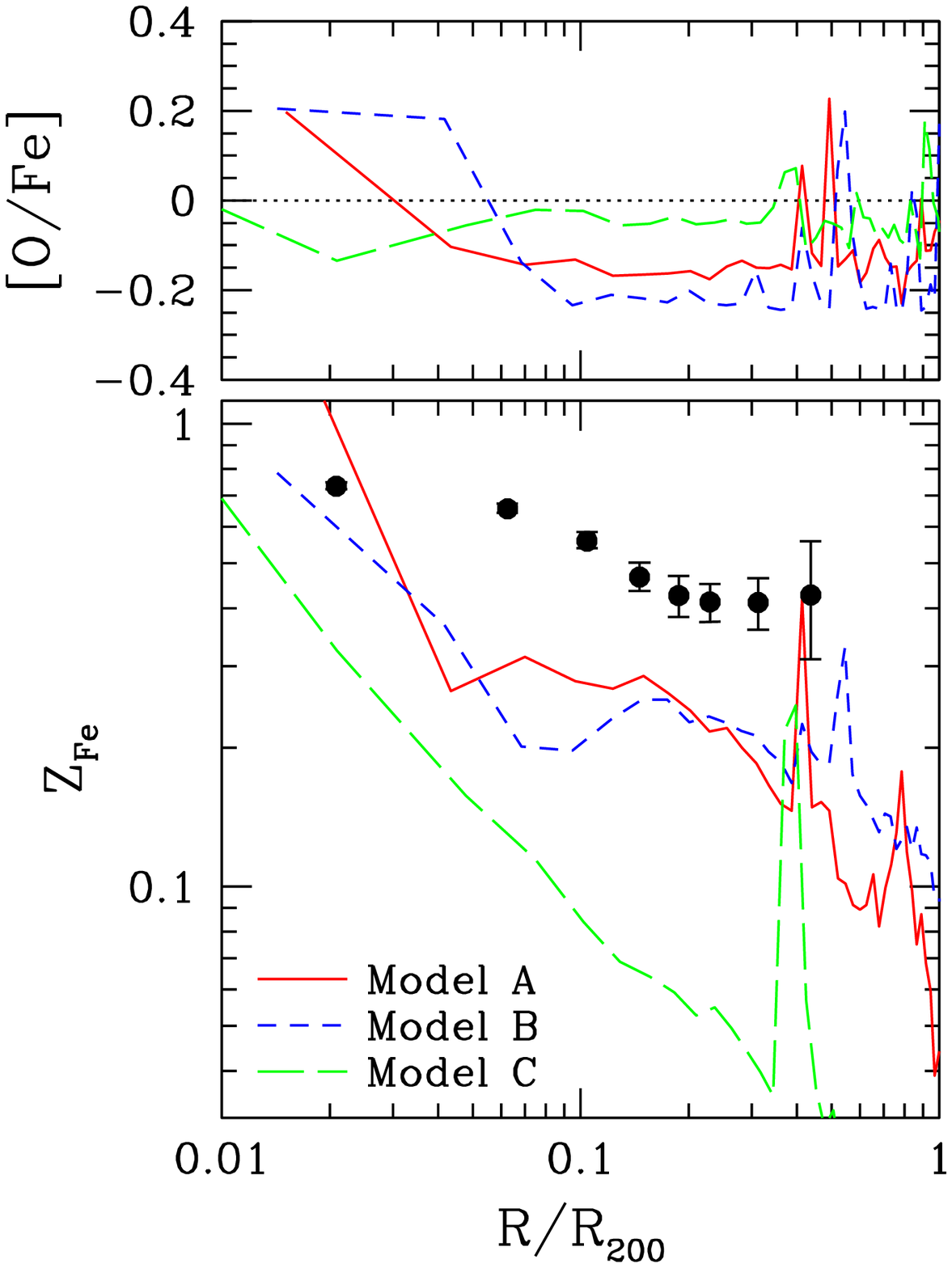,width=9.cm}
\hspace{-2truecm} 
\psfig{file=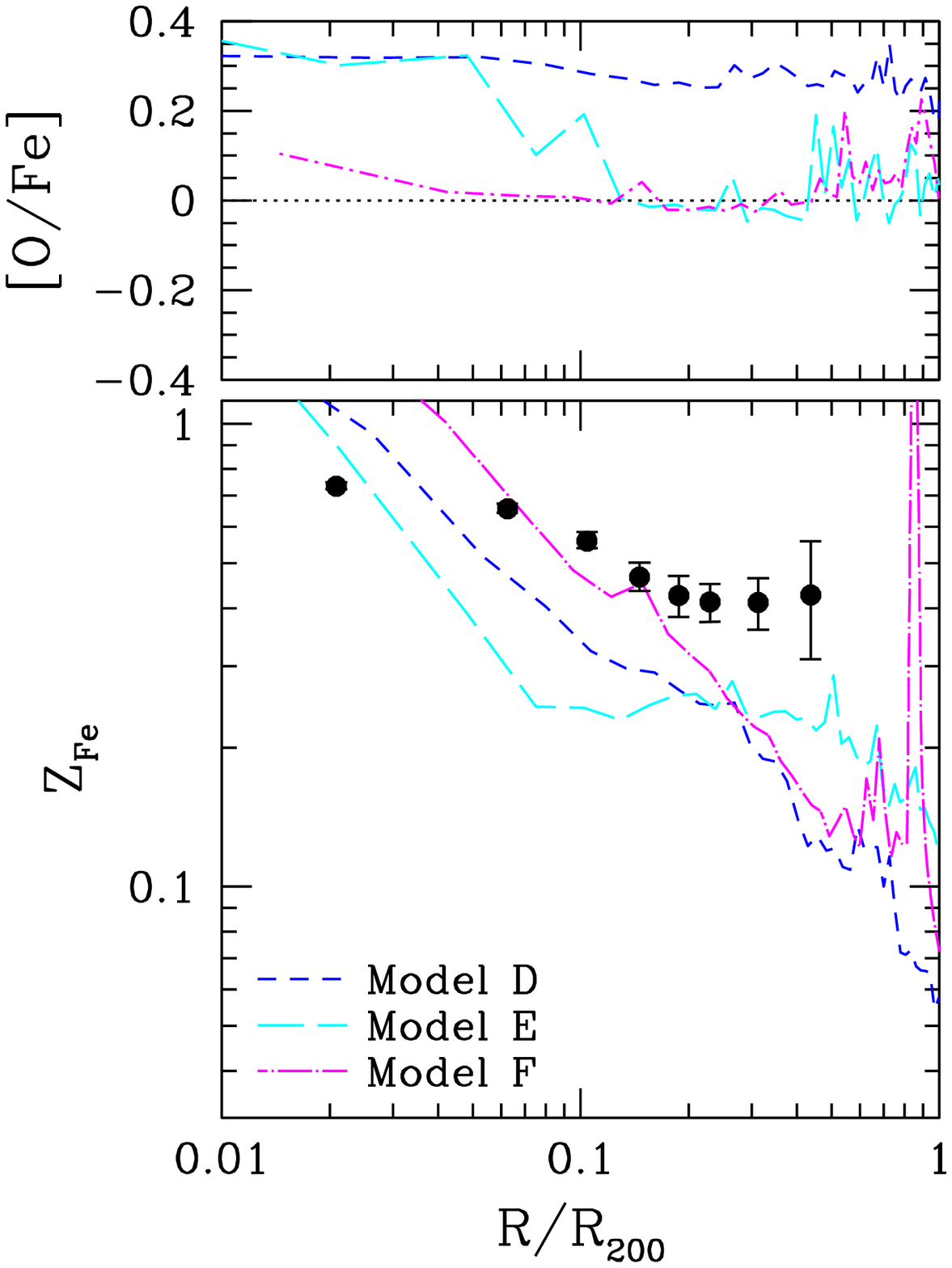,width=9.cm}
}}
\vspace{-2mm}
\caption{Profiles of the Fe abundance (lower panels) and of the [O/Fe]
  relative abundance (upper panels) for the models described in Table
  1. The points with errorbars in the lower panels show the observed
  average gradient of Fe abundance for cool--core clusters, as
  measured by De Grandi et al. (2003) from Beppo--SAX observations.}
\label{fi:metprof}
\vspace{-3mm}
\end{figure*}

\section{The simulation code}

Our simulations are based on an evolution of
GADGET\footnote{http://www.MPA-Garching.MPG.DE/gadget/} (Springel et
al. 2001), a parallel Tree+SPH code with fully adaptive
time--stepping. As a starting point, we used a version of GADGET,
kindly provided by V. Springel, which includes an entropy--conserving
integration scheme, radiative cooling, the effect of a uniform and
evolving UV background (Haardt \& Madau 1999), star formation from a
multiphase interstellar medium and a prescription for galactic winds
triggered by SN explosions (see SH03 for a detailed description). In
the original version of the code, the energy release and a global
metallicity was produced only by SNII under the
instantanous--recycling approximation (IRA).

The GADGET code has been suitably modified, so as to correctly include
the life--times of different stellar populations, to follow metal
production from both SNIa and II, while self--consistently introducing
the dependence of the cooling function on metallicity. A detailed
description of the implementation of these algorithms will be
presented in a forthcoming paper (Tornatore et al. 2004, in
preparation), while we provide here a short descriptions of the most
relevant features of the code.

\begin{table*}
\centering
\caption{Characteristics of the different runs. Col. 2: assumed IMF;
Col. 3: wind efficiency; Col. 4: energy injection efficiency; Col. 5:
average wind velocity (km s$^{-1}$); Col. 6: baryon fraction in stars
within the virial radius $R_{\rm vir}$; Cols. 7--10:
emission--weighted and mass--weighted ICM Iron and Oxygen abundances
within $R_{\rm vir}$ ($X/H$, solar units); Cols. 11--12: fraction of
total Iron and Oxygen masses, which are distributed in the diffuse
ICM. \newline $^*$ The wind velocity for Model F refers to redshift
range $z=0$--2, while it increases at higher $z$, reaching $v_W\simeq
1100$ km s$^{-1}$ at $z=10$.}
\begin{tabular}{lccccccccccc}
Run & IMF & $\eta$ & $\chi$ & $v_{W}$ & $f_*$ & $Z_{\rm Fe}^{ew}$ &
$Z_{\rm Fe}^{mw}$ & $Z_{\rm O}^{ew}$ & $Z_{\rm O}^{mw}$ & $f_{\rm Fe}^{ICM}$ & $f_O^{ICM}$\\ 
\hline 
A & S55 & 2 & 0.5 & 520 & 0.20 & 0.55 & 0.11 & 0.70 & 0.09 & 0.39 & 0.18\\ 
B & S55 & 2 & 1   & 380 & 0.26 & 0.23 & 0.14 & 0.20 & 0.09 & 0.34 & 0.13\\ 
C & S55 & 6 & 3   & 520 & 0.12 & 0.15 & 0.05 & 0.13 & 0.05 & 0.23 & 0.13\\
D & AY & 2 & 1 & 920 & 0.13 & 0.46 & 0.14 & 0.93 & 0.26 & 0.42 & 0.28\\
E & AY & 2 & 0.3 & 530 & 0.24 & 0.29 & 0.18 & 0.4 & 0.21 & 0.26 & 0.10\\
F & FBB& 2 & 1 & 600$^*$ & 0.15 & 0.54 & 0.17 & 0.59 & 0.18 & 0.41 & 0.24\\
\hline 
\end{tabular}
\label{t:runs}
\end{table*}

In order to maintain the general approach of the multiphase model by
SH03, we still treat under the IRA stars with masses $>20\,M_\odot$,
while accounting for the different life-times of stars of smaller mass
(Matteucci \& Padovani 1993). Within the stochastic approach to star
formation (SH03), each star particle is considered as a single stellar
population (SSP). For each SSP we compute the number of stars turning
into SNII and Ia at each time-step after its creation.

The SNIa are associated to binary systems whose components are in the
0.8--$8\,M_\odot$ mass range (Greggio \& Renzini 1983), while SNII
arise from stars with mass $>8\,M_\odot$ (cf. also Lia et al. 2002,
who adopt the lower mass threshold of $6M_{\odot}$ for SNII).  Besides
SNe, which release energy and metals, we also account for planetary
nebulae (PN). They contribute to metal production, but not to the
energy feedback, and are identified with those stars, not turning into
SNIa, in the mass range 0.8--8$\,M_\odot$.  We use the analytical
fitting formulas for stellar yields of SNIa, SNII and PNe as provided
by Recchi et al. (2001), and based on the original nucleosynthesis
computations of Nomoto et al. (1997, using their W7 model), Woosley \&
Weaver (1995) and Renzini \& Voli (1981). The formulation for the SNIa
rate has been calculated as in Matteucci \& Recchi (2001). Besides H
and He, the current version of the code follows the production of Fe,
O, C, Si, Mg, S, and can be easily modified to include other metal
species. Once produced by a star particle, metals are spread over the
same number of neighbours, 32, used for the SPH implementation, also
using the same kernel.  In this way, we find that 90 per cent of the
metals are distributed within a gas mass of $5.4\times 10^9
h^{-1}M_\odot$. We have verified that using a twice as large number of
neighbors to spread metals results in a twice as large gas mass for
metal mixing, while final results on the amount and distribution of
metals (see below) are left almost unchanged.  As for the energy
release, each SN is assumed to produce $10^{51}$ergs.  Instead of
assuming any specific value for the thermalization efficiency of the
energy released by SN, we prefer to dump all the energy to the
surrounding gas particles and leave to the simulation the computation
of the radiation losses. Since the physical processes determining the
actual SN efficiency are below the resolution scale of our
simulations, the rationale behind our choice is to leave to the
sub--grid multiphase model by SH03 establishing how much of this
energy enters in regulating the star formation process.
We normalize the IMFs in the mass range 0.1--100 $M_\odot$. Owing to
the uncertainty in modelling yields for very massive stars, we take
yields to be independent of mass above $40\,M_\odot$. While any
uncertainty in the yields of such massive stars has a negligible
effect for a Salpeter IMF (Salpeter 1955, S55 hereafter), their
accurate description (e.g. Thielemann et al. 1996; Heger \& Woosley
2002) is required when using a top--heavier IMF.  We note that our
scheme to distribute metals in the ICM does not include the effect of
diffusion. Lia et al. (2002) included the effect of diffusion driven
by SN blast waves (see Thornton et al. 1998) in their SPH simulations
with chemical enrichment. Although this effect is quite important to
describe the diffusion of metals within the interstellar medium, it is
likely to play a minor role on scales above the typical resolution
scale, $\sim 10\,h^{-1}$kpc, of the cluster simulations that we are
discussing here.

Our prescription to account for stellar evolution in the simulations
implies a substantial change of the multiphase ``effective model'' by
SH03. In this model, gas particles are assumed to have a cold neutral
and a hot ionized phase in pressure equilibrium, the former providing
the reservoir for star formation. We have modified the criterion,
dependent on local density and temperature, to establish the
relative amount of such two phases, so as to account for the gradual
SN energy release, which modifies the temperature of the hot phase and
the evaporation of the cold one. This means to include new energy
terms in Eq.(10) of SH03, which describes the evolution of the
internal energy of the hot--phase component, while making the cooling
function dependent on local metallicity. The metal--dependence of
cooling, that we introduce using the tables from Sutherland \& Dopita
(1993), also enters in determining the onset of the thermal
instability (eq.[22] of SH03) and the value of the density
threshold for star formation (eq.[23] of SH03).

SH03 also provided a phenomenological description for galactic winds,
which are triggered by SN energy release and whose strength is
regulated by two parameters. A first one gives the rate of gas
ejection by winds, according to the relation, $\dot M_W=\eta \dot
M_*$, where $\dot M_*$ is the star formation rate. The second one,
determines the fraction of SN energy that powers the winds, ${1\over
2}\dot M_W v_W^2= \chi \epsilon_{SN}\dot M_*$, where $\epsilon_{SN}$
is the energy feedback provided by the SN under IRA assumption for
each $M_\odot$ of stars formed. In our implementation of the winds, we
also account for the energy contributed from all the SN treated
without the IRA, namely SNII in the 8--20$M_\odot$ mass range and
SNIa.

It is worth noticing that the star formation and SN feedback scheme,
originally introduced by SH03, has been already demonstrated to
provide the correct fraction of baryons locked in stars, the correct
cosmic star formation history (Springel \& Hernquist 2003b), the
correct amount of neutral hydrogen in high column--density absorbing
systems at high redshift (Nagamine et al. 2003), and to reproduce the
basic X--ray scaling properties of galaxy clusters (Borgani et
al. 2003). As such, it represents a good starting point for a
simulation study of the ICM chemical enrichment.

\section{Results and discussion}
We use our chemodynamical version of GADGET to run simulations of the
Virgo--like cluster, which has been described by Tornatore et
al. (2003). This structure is a fairly relaxed halo with a virial mass
of $3.9\times 10^{14}M_\odot$ and emission--weighted temperature of
about 3 keV. It has been selected from a cosmological DM-only
simulation of a flat $\Lambda$CDM model, with $\Omega_m=0.3$,
$\Omega_{bar}=0.04$, $h=0.7$ and $\sigma_8=0.8$, within a box of 100
Mpc a side. Mass and force resolutions are increased in the Lagrangian
region surrounding the cluster, so that $m_{DM}=2.1\times 10^9M_\odot$
and $m_{gas}=3.2\times 10^8M_\odot$ for the mass of the DM and gas
particles, respectively. The Plummer--equivalent gravitational
softening is $\epsilon =5\,h^{-1}$ kpc fixed in physical units from
$z=0$ to $z=2$, while fixed in comoving units at earlier epochs.

In this Letter we focus on the effect of changing the main ingredients
which determine the content and distribution of metals in the ICM,
namely the IMF and the feedback strength. The choice of such
parameters for the different simulations, along with some results, are
summarized in Table \ref{t:runs}. Besides the standard IMF Salpeter
shape (S55), $dN/d\log{m}\propto m^{-x}$, with $x=1.35$, we also
consider the flatter IMF by Arimoto \& Yoshii (1987, AY hereafter),
with $x=0.95$, which provides a larger number of massive
stars. Different authors (e.g., Finoguenov et al. 2003, FBB hereafter;
Baumgartner et al. 2003, B03 hereafter) have suggested that an early
population of very massive, metal poor stars should be advocated to
account for the pattern of ICM abundances (cf. also Scannapieco,
Schneider \& Ferrara 2002). Therefore, we also consider an evolving
IMF with the shape proposed by Larson (1999), $dN/d\log{m}\propto
(1+m/m_s)^{-1.35}$. Following FBB, we allow $\log {m_s\ /M_\odot}$ to
linearly increase with redshift from $-0.35$ at $z=2$ up to 1 at
$z=10$, being constant at $z<2$.

As a first diagnostic for the ICM metal enrichment, we use the
profiles of Fe abundance. We compare in Figure \ref{fi:metprof} the
results of our simulations to those from the Beppo--SAX observations
of 12 cool--core clusters by De Grandi et al. (2003).  In the left
panels of Figure \ref{fi:metprof} we show the effect of changing the
strength of the feedback for the Salpeter IMF. Model A has a wind
speed of $v_W= 520$km s$^{-1}$. Model B assumes somewhat weaker winds,
while Model C assumes that the energy in winds is three times higher
than that provided by SN, with the same wind velocity as in Model A
once a three times larger amount of gas (i.e., $\eta=6$) is
ejected. As long as we assume that AGN activity in galaxies follows
star formation (e.g., Boyle \& Terlevich 1997, cf. also Cristiani et
al. 2003), such an extra energy powering the winds can be interpreted
as released by nuclear activity. All the models produce $Z_{\rm Fe}$
profiles (we define $Z_{\rm Fe}$ as the iron abundance by mass in
solar unit, $X_{\rm Fe}/X_{\rm Fe_\odot}$) which are steeper than
observed at $R\,\mincir 0.05R_{200}$.  At larger radii simulated
profiles for Models A and B have a shape similar to the observed ones,
although lower by about a factor two. An inspection of the age of
stars associated with the central cD reveals that they are younger
than those observed in real clusters, with signatures of significant
ongoing star formation down to $z=0$. This is also consistent with the
large amount of stars found in the simulated cluster (see Col. 6 of
Table \ref{t:runs}), witnessing that feedback is not strong enough to
inhibit recent star formation. The steep profiles of Fe abundance
suggest that a significant amount of Fe in central regions has been
produced quite recently, within an already formed cluster potential
well, thus making difficult for winds to transport metals far from the
star forming regions. This fact further confirms that a mechanism is
currently missing in simulations to diffuse metals and quench
star formation before most of the mass is accreted in the cluster
potential well. By increasing the wind efficiency, as in Model C, star
formation is heavely suppressed at high redshift. While this turns
into a more acceptable value of $f_*$, it has the unwelcome feature of
a too low ICM metallicity, with a quite steep gradient.

\begin{figure}
\vspace{-5mm}
\centerline{
\psfig{file=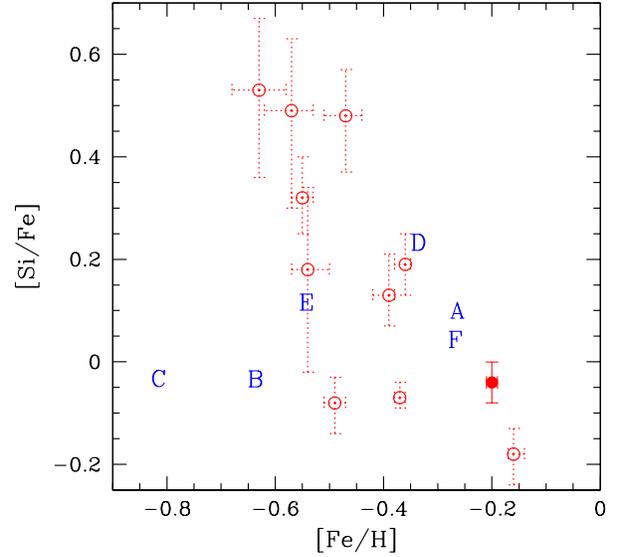,width=8.5cm}
}
\vspace{-2mm}
\caption{The comparison between the values
of Silicon-over-Iron ratio vs. Iron metallicity found by Baumgartner
et al. (2003) from their compilation of ASCA data (circles with
errorbars), and those measured for the different runs of our
Virgo--like clusters. Model labels are the same as in Table
\ref{t:runs}. The filled data point refers to the temperature bin
appropriate for our simulated cluster.}
\label{fi:sife}
\vspace{-3mm}
\end{figure}

Using for Model D the AY IMF, with the same choice of feedback
parameters as in Model A, provides a much larger energy release from
massive stars. Correspondingly, the wind velocity is about twice as
large and, probably, unrealistic. The resulting star
fraction is acceptable, but with a larger amount of Fe released into
the ICM, as a consequence of the higher IMF over most of the mass
range relevant for SNIa. Reducing the $\chi$ parameter in Model E, to
obtain wind velocity comparable to Model A, increases the resulting
star fraction and the total amount of produced Fe (see Table
1). However, the weaker winds cause a large Fe fraction to remain in
star--forming regions, being locked into stars instead of being
diffused.  Having a top--heavier IMF at high redshift (Model F)
has again the effect of increasing the feedback efficiency. As for the
Fe production and diffusion, its behaviour is similar to that of Model
D, thought with $\sim 30$ per cent more Fe due to the higher 
IMF over the mass range relevant for SNIa.

In the upper panels of Fig.\ref{fi:metprof} we show the profiles of
[O/Fe]. Since Oxygen is mostly produced by SNII, [O/Fe] is usually
considered as a diagnostic for the relative role played by the two SN
types. So far, robust measurements of the O abundance have been
realized with the XMM-NEWTON GRS, and are limited to the very
central regions of galaxy clusters. Gastaldello \& Molendi (2002) and
Matsushita et al. (2003) found [O/Fe]$\simeq -0.3$ around M87 out to
about 60 kpc, which corresponds to $\simeq 0.04\,R_{200}$ for our
simulated cluster. On larger scales, the difficult detection of the
OVIII emission line at 0.67 keV makes the determination of [O/Fe] more
uncertain, although values in the range between $-0.2$ and 0.2 seem to
be preferred (see Renzini 1997 for a discussion).
The negative gradients of our [O/Fe] profiles are the signature of
recent star formation episodes in the central region. This arises
because SNII already had time to explode, while SNIa, which are
generated by longer--living stars, still have to release Iron. As a
consequence, the ICM for Models A and B turns out to be over--abundant
in Oxygen, in the central regions, with respect to what observed for
M87, while being consistent with data on larger scales. Assuming an AY
IMF provides a relatively larger number of SNII, thus exacerbating the
problem of Oxygen overabundance.

As for the overall metallicity, we obtain emission--weighted $Z_{\rm
Fe}$ values in the range 0.25--0.55, with the only exception of Model
C, which produces a lower Fe abundance.  It is worth noticing that,
although Models A and F produces the highest $Z_{\rm Fe}^{ew}$ values
(see Table 1), they have different mass--weighted Fe abundances. In
particular, the high $Z_{\rm Fe}^{ew}$ of Model A corresponds to a
fairly small mass--weighted value, the large difference being due to
the steep profile of the Fe abundance. In their compilation of ASCA
data, B03 find that $Z_{\rm Fe}$ tends to decrease with cluster
temperature, with $Z_{\rm Fe}\simeq 0.7$ for $T\simeq 3$ keV, down to
$Z_{\rm Fe}\simeq 0.3$ at $T\simeq 10$ keV (cf. also Renzini 2003, who
claims a roughly constant $Z_{\rm Fe}\sim 0.3$ over the same
temperature range). Therefore, for our Virgo--like cluster we should
expect a Fe enrichment which is sensibly larger than found in
simulations. However, we note that the procedure of stacking X--ray
spectra of clusters with similar temperature, as followed by B03,
corresponds to assuming that all clusters having the same $T$ also
have the same metallicity, and, therefore, that the fairly large
scatter observed is only due to measurement errors. Since some amount
of intrinsic scatter is anyway expected, it is difficult to assess,
with only one cluster simulated, how discrepant are our $Z_{\rm Fe}$
values with respect to the observed ones. Another feature of the
simulated Fe production is that the ICM contains between about $25$
and 40 per cent of the total amount of produced Fe, at odds with
observational indications for $M_{\rm Fe,ICM}\,/M_{\rm Fe,*}\sim 2$
(Renzini 2003). This further suggests that a mechanism is still
lacking in simulations to transport metals away from star--forming
regions.

Unlike Oxygen, which is difficult to detect in the X--ray spectra of
the hot ICM, Silicon is now detected for a fair number of clusters.
Since a sizeable fraction of Si is produced also by SNIa, depending on
the choice for the IMF, it could be a useful tool to investigate the
relative contribution to metal enrichment from the two SNe types
(e.g. Lowenstein 2003).  In Figure 2 we show [Si/Fe] vs. [Fe/H] for
our simulations, compared to the observational values from
B03. Several of our runs produce a relative [Si/Fe] ratio which falls
on the correlation indicated by data, although the observed absolute
Fe abundance for $\simeq 3$ keV clusters (marked by the filled circle)
is significantly larger than that produced by several of our models.
We note that the models better approximating observational data are A
and F, which, besides providing a realistic emission--weighted Fe
abundance, also produce an acceptable amount of Si.

As a word of caution, we note that the resolution of the simulations
presented here could prevent the treatment of some physical effects,
which may change final results. For instance, effects of dynamical
stripping may play a significant role in removing metal--rich gas from
cluster galaxies and, therefore, to enrich the diffuse ICM (e.g.,
Gnedin 1998; Toniazzo \& Schindler 2001). While tidal stripping should
be well represented in our simulations, a proper treatment of
ram--pressure stripping would require a substantially higher
resolution in an SPH simulation. Aguirre et al. (2001) claim that
ram--pressure stripping from galaxies of mass $>3\times
10^{10}M_\odot$ accounts for a small fraction of the ICM metal
content. Furthermore, Renzini (2003) argues that, if ram--pressure
stripping were efficient, then clusters with larger velocity
dispersion should have a relatively higher metallicity, a trend which
is not observed.

Furthermore, in order to have reliable estimates of the ICM metal
enrichment, we have to make sure that our simulation has high enough
resolution to provide a correct representation of the star formation
history (see also Tornatore et al. 2003). We postpone a detailed study
of numerical convergence of star formation and metal production within
clusters in a forthcoming paper.

\section{Conclusions}
We have presented results from hydrodynamical simulations of the ICM,
realized with a chemodynamical version of the GADGET code. We used a
version of this code that, besides a fairly advanced treatment of star
formation and feedback from galactic winds (Springel \& Hernquist
2003a, SH03), also correctly accounts for life--times of different
stellar populations, as well as for metal and energy release from SNIa
and II (Tornatore et al. 2004, in preparation). By simulating one
single cluster, having temperature of about 3 keV, we looked at the
effect of feedback strength and IMF on the resulting metal
production. Our main results can be summarized as follows.\\

\noindent
{\bf(a)} Among the considered models, using a variable IMF (Model F)
provides an acceptable amount of Fe mass, as well as [Si/Fe] and
[O/Fe] ratios, in fair agreement with observations.  Using a Salpeter
(1955) IMF turns into a lower amount of Fe mass and to supersolar
values of [O/Fe] in the central cluster region, at variance with
observations. An Arimoto--Yoshii (1987) IMF provides an even larger
Oxygen abundance, as a consequence of the larger number of SNII. 

\noindent
{\bf (b)} Gradients of the Fe abundance are always steeper than
observed in central cluster regions, $R\mincir 0.1R_{200}$, while the
$Z_{\rm Fe}$ profiles on larger scales fall below the observed
ones, by an amount which depends on the feedback strength and on
the assumed IMF. 

\noindent
{\bf (c)} Besides the effect of the IMF, the oversolar value of the
[O/Fe] ratio found in central regions for several models could be due
to the presence of a significant recent star formation in central
cluster regions. Because of such an excess of recent star formation,
not shown by observational data, the stars in our simulations lock
a large fraction of metals (see Table 1).

Our analysis confirms that the observed pattern of metal enrichment of
the ICM is deeply connected to the past history of star formation in
clusters and to the feedback scheme that should release energy and
heavy elements into the diffuse medium. The fairly steep gradients of
the Fe abundance and the signature of recent star formation
consistently indicate that our simulations are missing a feedback
mechanism, which should quench star formation and spread metals at an
early enough epoch, when the proto--cluster potential well is still
shallow enough not to retain the produced heavy elements. We verified
that this feedback cannot be easily implemented by resorting to SNII
and Ia. Neither increasing the strength of the feedback nor changing
the IMF helps in restoring a better agreement with the observed
profile of the Fe abundance. In our opinion this is a non--trivial
result, since it has been obtained in the framework a rather advanced
scheme for star formation and SN feedback (SH03), which is otherwise
quite successful at accounting for the pattern of cosmic star
formation (Springel \& Herquist 2003b).  In turn, this suggests that
other energy sources, such as AGN, should be called into play to
regulate the cooling structure of the ICM and determine its chemical
composition.

\section*{Acknowledgments.}
We are greatly indebted to Volker Springel for having provided us with
the non--public version of GADGET, and for his continuous advices on
the code whereabouts. We acknowledge useful discussions with Francesco
Calura, Cristina Chiappini, Sabrina De Grandi, Alexis Finoguenov and
Laura Portinari. We thank the referee Frazer Pearce for his useful and
clever comments. The simulations have been realized using the IBM-SP4
machine at the ``Centro Interuniversitario del Nord-Est per il Calcolo
Elettronico'' (CINECA, Bologna), with CPU time assigned thanks to an
INAF--CINECA grant.

\end{document}